\documentclass[12pt]{article}
\usepackage{amssymb}
\usepackage{amsmath,graphicx,amsfonts,epsfig,eucal}
\usepackage{amsfonts}
\usepackage{color}
\usepackage{amsthm}

\usepackage[utf8]{inputenc}
\usepackage[english]{babel}

\newtheorem{prooft}{Proof of Theorem}

\newtheorem{thm}{Theorem}

\newtheorem{prop}{Proposition}
\newtheorem{defn}{Definition}
\newtheorem{rem}{Remark}

\begin{document}
\title{Elasticity of substitution and general model of economic growth}
\author{Constantin Chilarescu\footnote{e-mail: constantin.chilarescu@univ-lille.fr}}
\date{}
\maketitle
\centerline{\it \it University of Lille, France and West University of Timisoara, Romania}
\maketitle
\begin{abstract}
The main purpose of this paper is to generalize some recent results obtained by Chilarescu and Manuel Gomez. Essentially, we are trying to study the effect of elasticity of substitution on the parameters of economic growth, based on its two possible values - lower and higher than one. We show that a higher elasticity of substitution increases per capita income, the relative share of physical capital, the common growth rate and the share of human capital allocated to the production sector, and this property is not affected by the position of the elasticity of substitution - below or above one.
\end{abstract}
\noindent \textit{\\ Keywords:} Hamiltonian function; balanced growth;
elasticity of substitution; normalized functions.
\noindent \textit{\\ JEL\ Classifications:} C61, C62, E21, O41.
\date{}
\maketitle
% ----------------------------------------------------------------
\section{Introduction}
A recent paper of Manuel Gomez $(2024)$ published by this journal, studied the effect of the elasticity of substitution on the long-run growth and the convergence speed in the one sector endogenous growth model, using a general production function that encompasses the most commonly used specifications as particular cases. In fact, this paper is the latest essay in a long attempt on this subject, almost all devoted to the one-sector endogenous growth (see the paper of this author $2016 - 2024$).

A recent paper of Chilarescu $(2024)$, proposes an improvement of some recent results on the impact of the elasticity of  substitution on economic growth. First of all, the author chose an economy characterized by two sectors, both modelled by a production function in which physical and human capital are simultaneously present, while other papers used an economy characterized by a single sector. To do this, the author has modified the model developed by Rebelo $(1991)$, considering that the two sectors of the economy are described by two $CES$ production functions, distinct regarding the parameters of efficiency and distribution, but having a common substitution coefficient. He also assumes that the elasticity of substitution $\sigma$ can only take values greater than one, that is to say that the values of the substitution parameter $\psi$ must be in the interval $(0, 1)$. Under these hypothesis, the author proves that an increase in the elasticity of substitution will produce an increase of: per-capita income, relative share of physical capital, common growth rate and the fraction of the human capital allocated to the good sector. These results are the same as those obtained for the case of a one-sector model.

The problem there is that the restriction $\sigma \geq 1$ is absolutely necessary in order to have a long-run growth rate positive (see the paper of Palivos and Karagiannis, $2007$ and the paper of Manuel Gomez, $2020$, $2021$, $2023$ and $2024$). An interesting result was recently obtained by Ozkaya $(2022)$, who claim that if $\sigma > 1$, then  the effect of the elasticity of substitution on the share of capital depends on the ratio of minimum marginal products and if $\sigma < 1$, then the effect of the elasticity of substitution on the share of capital depends on the ratio of maximum marginal products. Of course, it will also be very important to know what happens to the growth parameters when $\sigma$ takes values not only greater than one, but also when its values are in the interval $(0.1)$.

In this paper, we try to generalize the latest results of Chilarescu by considering a general model of endogenous growth of Bond et al. $(1996)$ type, that is to say where the two sectors are modelled by two completely distinct $CES$ production functions. The methodology used in the paper is that developed by de La Grandville $(1989)$ and Klump and de La Grandville $(2000)$, followed later by a considerable number of papers. We try to highlight the effect of the elasticity of substitution, according to its positive values, lower or higher than one, on the main parameters that characterize economic growth.

The paper is organized as follows. The next section presents the model with two distinct $CES$ production functions and prove the existence and uniqueness of the steady-state equilibrium. In the third section we prove the existence of a unique balanced growth path and then we examine the local stability of this steady-state. The fourth section studies the effect of the elasticity of substitution on long-run equilibrium and on steady state and the final section gives some conclusions.
\section{The general two-sector model}
The two sectors considered in this paper are the good sector that produces consumable and gross investment in physical capital, and the education sector that produces human capital, both of them under conditions
of constant returns to scale. Without loss of generality, we suppose that the economy is populated by a large and constant number of identical agents, normalized to one, who seek to maximize their utility function.
The good sector combines a fraction $v$ of the stock of physical capital with a fraction $u$ of the stock of human capital. In his recent paper, Chilarescu also assume that the fraction $v$ of physical capital allocate to the good sector is greater than the fraction $1-v$ allocated to the educational sector. The same assumption is also considered for the fraction $u$ of human capital, $i.e.$, $u > 1-u$. He argues these statements based on the papers of Lucas $(1988)$, where $u$ at steady-state was $0.82$ and the transitional values of $u$ lie in the interval $(0.50; 0.85)$, according with the papers of Boucekkine and Ruiz-Tamarit, $2008$ and Chilarescu $2011$. Although these assumptions seem reasonable, in this paper we do not use them.
The output of the educational sector increases the supply of effective labor units by adding to the existing stock of human capital, that will then be used in both sectors. The educational sector combines the remaining fractions of stocks.
Our model is characterized by the following optimization problem.
\begin {defn}
The set of paths $\left\{k, h, c, u, v\right\}$ is called an optimal
solution if it solves the following optimization problem:
\begin {equation}\label{eqof}
V_0 = \max
\int\limits_0^{\infty}\frac{c^{1-\varepsilon}-1}{1-\varepsilon}e^{-\rho t}dt,
\end {equation}
\noindent subject to
\begin {equation}\label{eqres}
\left\{
 \begin{array}{lll}
\dot{k} = y_1- c-\delta_{k} k,\\\\
\dot{h} =y_2-\delta_{h} h,\\\\
k_0 = k(0),\;h_0 = h(0),\\
  \end{array}
  \right.
\end {equation}
where
$$y_1=A_1\left[\alpha_1\,\left(kv\right)^{\psi_1}+\left(1-\alpha_1
\right)\left(hu\right)^{\psi_1}\right]^{\frac{1}{\psi_1}},$$ $$y_2=A_2\left\{\alpha_2\,\left[k\left(1-v\right)\right]^{
\psi_2}+\left(1-\alpha_2\right)\left[h\left(1-u\right)\right]
^{\psi_2}\right\}^{\frac{1}{\psi_2}}.$$
\end {defn}
\noindent In this model $h$ represents the human capital
or the skill level and $k$ is physical capital per labor unit. $\alpha_1 \in (0,1)$ and $\alpha_2 \in (0, 1)$ are distribution parameters,
$\psi_{i} = \frac{\sigma_{i}-1}{\sigma_{i}},\;\psi_{i} < 1,\; i = 1, 2$ are the substitution parameters,
$\rho$ is the rate of time preference, $\delta_{k}$ and $\delta_{h}$ are the depreciation rates,
$A_1$ and $A_2$ are efficiency parameters, and $c \geq 0$ is
the real per-capita consumption. $\varepsilon^{-1}$ represents the
constant elasticity of intertemporal substitution and $k_0$ and $h_0$ are given values.
To solve the problem \eqref{eqof} subject to  \eqref{eqres}, we define the
Hamiltonian function:
\begin{equation}\label{eqham}
H = \frac{c^{1-\varepsilon}-1}{1-\varepsilon} + \left(y_1 - c-\delta_{k} k\right)\lambda + \left(y_2-\delta_{h} h\right)\mu.
\end{equation}
\noindent The boundary conditions include initial conditions $\bar{h},
\;\bar{k}$ and transversality condition:
$$\lim\limits_{t\rightarrow\infty} e^{-\rho t}\lambda(t)
k(t) = 0 \;\;\mbox{and}\;\; \lim\limits_{t\rightarrow\infty}
e^{-\rho t}\mu(t) h(t) = 0.$$
In our model there are three control variables, $c$, $v$ and $u$, and two
state variables, $k$ and $h$. Differentiating the Hamiltonian function
with respect to $u$ and then with respect to $v$, we obtain the
following relations:
\begin{equation}\label{eqc}
c^{-\varepsilon} = \lambda,
\end{equation}
\begin{equation}\label{eqderu}
\frac{A_2\left[\bar{P}_2\right]^{\frac{1}{\psi_2}-1}\left(1-\alpha_2 \right)
\left[h\left(1-u\right)\right]^{\psi_2}\mu}{1-u}=\frac {A_1\left[\bar{P}_1\right]^{\frac{1}{\psi_1}-1}\left(1-\alpha_1\right)\left(hu\right)^{\psi_1}\lambda}{u},
\end{equation}
\begin{equation}\label{eqderv}
\frac {A_2\left[\bar{P}_2\right]^{\frac{1}{\psi_2}-1}\alpha_2\left[k\left(1-v\right)\right]^{\psi_2}\mu}{1-v
}=\frac{A_1\left[\bar{P}_1\right]^{\frac{1}{\psi_1}-1}\alpha_1\,\left(kv\right)^{\psi_1}\lambda}{v},
\end{equation}
where
$$\bar{P}_1=\alpha_1\left(kv\right)^{\psi_1}+\left(1-\alpha_1
\right)\left(hu\right)^{\psi_1},\bar{P}_2=\alpha_2\left[k\left(1-v\right)\right]^{
\psi_2}+\left(1-\alpha_2\right)\left[h\left(1-u\right)\right]^{\psi_2}$$
From equations \eqref{eqderu} and \eqref{eqderv}, we get:
\begin{equation}\label{eqsr}
\frac{\alpha_1(1-\alpha_2)}{\alpha_2(1-\alpha_1)}\left[\frac{kv}{hu}\right]^{\psi_1-\psi_2}=
\left[\frac{v(1-u)}{u(1-v)}\right]^{1-\psi_2}\;\Leftrightarrow\;\theta w^{\psi_1-\psi_2}=\tau^{1-\psi_2},
\end{equation}
and
\begin{equation}\label{eqlm}
\frac{\mu}{\lambda} = \frac{A_1\alpha_1}{A_2\alpha_2\theta}\frac{P_1^{\frac{1}{\psi_1}-1}}{P_2^{\frac{1}{\psi_2}-1}},\;
\theta=\frac{\alpha_1(1-\alpha_2)}{\alpha_2(1-\alpha_1)},\;\tau=\frac{v(1-u)}{u(1-v)},
\end{equation}
where $z=\frac{k}{h}$, $w=\frac{v}{u}z$ and
\begin{equation}\label{eqp1p2}
P_1=\alpha_1 w^{\psi_1}+1-\alpha_1,\; P_2=\alpha_2\theta^{-\frac{\psi_2}{1-\psi_2}}
w^{\frac{\psi_2(1-\psi_1)}{1-\psi_2}}+1-\alpha_2.
\end{equation}
Differentiating the Hamiltonian function with respect to $k$ and then with respect to $h$, we obtain
the differential equations describing the trajectories of the dual variables.
\begin{equation}\label{eqsysb1}
\frac{\dot{\lambda}}{\lambda}=\rho+\delta_{k}-\alpha_1A_1w^{\psi_1-1}P_1^{\frac{1}{\psi_1}-1},\;
\frac{\dot{\mu}}{\mu}=\rho+\delta_{h}-\left(1-\alpha_2\right)A_2 P_2^{\frac{1}{\psi_2}-1}.
\end{equation}
Differentiating with respect to time the equations \eqref{eqsr} and \eqref{eqlm} we obtain the following differential equations:
\begin{equation}\label{eqdifuv1}
\frac{G_1}{1-u}\frac{\dot{u}}{u}-\frac{G_2}{1-v}\frac{\dot{v}}{v}=
(\psi_1-\psi_2)\left(D+\frac{c}{k}\right),
\end{equation}
\begin{equation}\label{eqdifuv2}
\frac{\dot{u}}{u}-\frac{\dot{v}}{v}=-\left(D+\frac{c}{k}\right)-\frac{QP}{\left(1-\psi_1\right)T},
\end{equation}
where
$$\frac{\dot{k}}{k}-\frac{\dot{h}}{h}=-\left(D+\frac{c}{k}\right),
\;\frac{\dot{\mu}}{\mu}-\frac{\dot{\lambda}}{\lambda}=P,$$
$$D=A_2(1-u)P_2^{\frac{1}{\psi_2}}
-A_1vw^{-1}P_1^{\frac{1}{\psi_1}}+\delta_k-\delta_h,$$
$$P=\alpha_1A_1w^{\psi_1-1}P_1^{\frac{1}{\psi_1}-1}-\left(1-\alpha_2\right)A_2P_2^{\frac{1}{\psi_2}-1}
-\left(\delta_k-\delta_h\right),$$
$$T=\alpha_1\left(1-\alpha_2\right)w^{\psi_1}
-\alpha_2\left(1-\alpha_1\right)\theta^{\frac{-\psi_2}{1-\psi_2}}
w^{\frac{\psi_2(1-\psi_1)}{1-\psi_2}},$$
$$G_1=(\psi_1-\psi_2)u+1-\psi_1,\;G_2=(\psi_1-\psi_2)v+1-\psi_1,$$
$$Q=P_1P_2,\;R=\left(1-\psi_1\right)\left(1-\psi_2\right)T.$$
Now, we can close the system and write down the final form of the five differential equations,
\begin{equation}\label{eqsysb}
\left\{
  \begin{array}{llllll}
\frac{\dot{k}}{k} = A_1v{w^{-1}}P_1^{\frac{1}{\psi_1}} - \frac{c}{k}-\delta_{k},\\\\
\frac{\dot{h}}{h} =A_2P_2^{\frac{1}{\psi_2}}(1-u) -\delta_{h},\\\\
\frac{\dot{c}}{c} = -\frac{\rho+\delta_{k}}{\varepsilon}+\frac{\alpha_1A_1w^{\psi_1-1}}{\varepsilon}P_1^{\frac{1}{\psi_1}-1},\\\\
\frac{\dot{u}}{u}=\left[D+\frac{c}{k}+\frac{Q G_2P}{R}\right]\frac{1-u}{u-v},\\\\
\frac{\dot{v}}{v}=\left[D+\frac{c}{k}+\frac{Q G_1P}{R}\right]\frac{1-v}{u-v}.\\
\end{array}
  \right.
\end{equation}
\section{The balanced growth path}
To simplify the notation we denote by $r_x$ the growth
rate of variable $x$, by $x^*$ its value along the balanced growth
path $(t \geq t_*)$, and by $x_*$ its value for $t = t_*$. The following
proposition gives our first main result.
\begin{prop}\label{prop1}
If $r_u = r_v = 0$ for all $t \geq t_* $, then the above system reaches the unique balanced growth path and the following statements are valid
\begin{enumerate}
\item[i.] $r_{k^*} = r_{h^*} = r_{c^*}  = r_{y^*_1} = r_{y^*_2} = r_*$, and
\begin{equation}\label{eqsolr}
r_*=\frac{1}{\varepsilon}\left[\alpha_1A_1w^{\psi_1-1}P_1^{\frac{1}{\psi_1}-1}
-\rho-\delta_{k}\right]
\end{equation}
\item [ii.] $w_{*}$ is the unique solution of the equation
\begin{equation}\label{eqsolz}
P(w) =\alpha_1A_1w^{\psi_1-1}P_1^{\frac{1}{\psi_1}-1}-\left(1-\alpha_2\right)A_2P_2^{\frac{1}{\psi_2}-1}
-\left(\delta_k-\delta_h\right)=0,
\end{equation}
\item[iii.] $u_* \in [0, 1]$, $v_* \in [0, 1]$, $\tau_0=w^{\frac{\psi_1-\psi_2}{1-\psi_2}}\theta^{\frac{1}{1-\psi_2}}$ and
\begin{equation}\label{eqsolu}
u_* = 1-\frac{r_*+\delta_h}{A_2P^{\frac{1}{\psi_2}}_2},\;v_* = \frac{\tau_0 u_*}{1+\left(\tau_0-1\right)u_*},
\end{equation}
\item[iv.]
\begin{equation}\label{eqsolc/k}
q_*=\frac{c_*}{k_*} = \frac{1}{\varepsilon} \left[A_1w^{-1}P_{\varepsilon}P_1 ^{\frac{1}{\psi_1}-1}+\rho-\delta_k\left(\varepsilon-1\right)\right],
\end{equation}
where $$P_{\varepsilon}=\alpha_1(\varepsilon v_*-1)w^{\psi_1}+\varepsilon(1-\alpha_1)v_*.$$
\end{enumerate}
\end{prop}
\noindent {\bf Proof of Proposition 1.}
If $r_u=r_v =0$ for all $t \geq t_*$ and since $P_1P_2 \neq 0$ for all $t \geq t_*$, from equations \eqref{eqdifuv1} and \eqref{eqdifuv2}, it immediately follows that $r_{k^*} = r_{h^*}$ and $r_{\mu^*}= r_{\lambda^*}$.
Of course, because we have $r_{k_*} = r_{h_*}$, we also must have $r_{y_{1*}}=r_{y_{2*}}=r_{k_*}$. This statement is obtained simply by log-differentiating the two production functions. From the two equations of \eqref{eqsysb1} we obtain \eqref{eqsolz} and $w_*$ will be the unique solution of this equation. (It is just a simply exercise to prove that this equation has a unique solution. Indeed, $\lim\limits_{w\rightarrow 0}P(w) = +\infty$, $\lim\limits_{w\rightarrow +\infty}P(w) = -\infty$ and $P^{\prime} < 0$ for all $z > 0$. Consequently, the function $P$ is a strictly decreasing function on $(0, \infty)$ and therefore there exists a unique solution of the equation $P(w) = 0.$) Substituting now $w_*$ into the third equation of the system \eqref{eqsysb} we get the common growth rate of the economy given by \eqref{eqsolr}. Substituting $r_*$ from equation \eqref{eqsolr} into the second equation of  the system \eqref{eqsysb} we obtain \eqref{eqsolu}.
We can now substitute this result into the first equation of the system \eqref{eqsysb} to obtain the steady state value for the variable $\frac{c}{k}$ given in equation \eqref{eqsolc/k}. Unfortunately we cannot prove that $u_* \in (0, 1)$. Nevertheless, all the numerical simulations we made, confirms that it lies inside the interval $(0, 1)$. Under the hypotheses $u_*\in (0, 1)$ it immediately follows that $v_*\in (0, 1)$ and thus we may claim that the proof is completed.$\square$

We have now to verify if the steady-state found above, satisfies the transversality conditions. Equivalently, the two conditions are true if the following limits are both negative, that is
$l_1=\lim\limits_{t\rightarrow\infty}\left(\frac{\dot{k}}{k}+\frac{\dot{\lambda}}{\lambda}-\rho\right)<0$ and $l_2=\lim\limits_{t\rightarrow\infty}\left(\frac{\dot{h}}{h}+\frac{\dot{\mu}}{\mu}-\rho\right)<0.$
Substituting the corresponding elements from the system \eqref{eqsysb} and passing to the limit for $t\rightarrow\infty$ we obtain for both limits the following result:
$l_1= l_2 = -\left[\rho+\left(\varepsilon-1\right)r_*\right].$
This limit is obviously negative, for all $\varepsilon > 1$, that is if the inverse of the elasticity of intertemporal substitution is greater than one. Without loss of generality, everywhere in this paper we assume that this inequality holds and therefore, since both transversality conditions are true, we may conclude that the solution obtained for the steady-state, is the unique optimal solution.
\subsection{Stability analysis}
In this section we examine the local stability of the steady-state found in the previous section. In order to do that, we first reduce the system in order to obtain a more easier dimensional one.  We therefore try to determine some transformed variables that are stationary at the $BGP$.
The dynamics of the economy around the steady-state, in terms of the new variables $z=\frac{k}{h}$, $q = c/k$, $u$ and $v$ - which are constant at the balanced growth path - is described by the following system of equations:
\begin{equation}\label{eqsysbgp}
\left\{
  \begin{array}{llll}
\dot{z}= -\left\{D(z,u,v) + q\right\}z\\\\
\dot{q} = \left\{q-\frac{A_1H(z,u,v)}{\varepsilon} -\frac{\rho-(\varepsilon-1)\delta_{k}}{\varepsilon}\right\}q,\\\\
\dot{u}=\left[D(z,u,v)+q+\frac{G_2(v)Q(z,u,v)P(z,u,v)}{R(z,u,v)}\right]\frac{u(1-u)}{u-v},\\\\
\dot{v}=\left[D(z,u,v)+q+\frac{G_1(u) Q(z,u,v)P(z,u,v)}{R(z,u,v)}\right]\frac{v(1-v)}{u-v},\\
\end{array}
  \right.
\end{equation}
where $$H(z,u,v) = \left[P_1(z,u,v)\right]^{\frac{1}{\psi_1}-1}P_{\varepsilon}(z,u,v),\;Q(z,u,v)=P_1(z,u,v)P_2(z,u,v).$$
The Jacobian matrix of system \eqref{eqsysbgp} evaluated at the
steady state is defined as follows
\begin{equation}\label{eqJst}
J_*=\left[J_{ij}\right](x_*),\;i,j = 1, 2, 3, 4,\;\;x_*=\left(z_*,q_*,u_*, v*\right).
\end{equation}
$$J_{11}=-\frac{\partial D}{\partial z}z,\;J_{12}=-z,\;J_{13}=-\frac{\partial D}{\partial u}z,\;
J_{14}=-\frac{\partial D}{\partial v}z,$$
$$J_{21}=-\frac{A_1}{\varepsilon}\frac{\partial H}{\partial z}q,\; J_{22}=q,
J_{23}=-\frac{A_1}{\varepsilon}\frac{\partial H}{\partial u}q,\;J_{24}=-\frac{A_1}{\varepsilon}\frac{\partial H}{\partial v}q,$$
$$J_{31}=\left[\frac{\partial D}{\partial z}-\frac{G_2Q}{R}\frac{\partial P}{\partial z}\right]\frac{u(1-u)}{u-v},\;J_{32}=\frac{u(1-u)}{u-v},$$
$$J_{33}=\left[\frac{\partial D}{\partial u}-\frac{G_2Q}{R}\frac{\partial P}{\partial u}\right]\frac{u(1-u)}{u-v},\;J_{34}=\left[\frac{\partial D}{\partial v}-\frac{G_2Q}{R}\frac{\partial P}{\partial v}\right]\frac{u(1-u)}{u-v},$$
$$J_{41}=\left[\frac{\partial D}{\partial z}-\frac{G_1Q}{R}\frac{\partial P}{\partial z}\right]\frac{v(1-v)}{u-v},\;J_{42}=\frac{v(1-v)}{u-v},$$
$$J_{43}=\left[\frac{\partial D}{\partial u}-\frac{G_1Q}{R}\frac{\partial P}{\partial u}\right]\frac{v(1-v)}{u-v},\;J_{44}=\left[\frac{\partial D}{\partial v}-\frac{G_1Q}{R}\frac{\partial P}{\partial v}\right]\frac{v(1-v)}{u-v}.$$
All these values are computed into the steady-state point $\left(z_*,q_*,u_*, v*\right).$
Unfortunately, the results obtained for the Jacobian matrix do not allow an incontestable interpretation of the signs of these eigenvalues. Nevertheless, all the numerical simulations confirm that at least one eigenvalue is negative and therefore, we can claim that there exists a unique optimal steady-state equilibrium that is saddle-path stable. The benchmark values chosen for the economy we consider are the following:
$A_1 = 1.05, A_2 = 0.20, \alpha_1 = 0.6, \alpha_2 = 0.8, \delta_k = 0.06, \delta_h = 0.05, \varepsilon = 2, \rho = 0.06$ and the results of the numerical simulations for the five distinct cases, are ($EV$ signify eigenvalues):
\begin{itemize}
  \item[1] $\psi_1 = 0.25$, $\psi_2 = -0.10$, $\sigma_1 =1.33$, $\sigma_2 = 0.91$,
  $z_* = 10.73, u_* = 0.882, v_*= 0.866, q_*= 0.240$,
  $EV=\left[0.0014; 0.173; 12.963; -12.788\right]$,
  \item[2] $\psi_1 = -0.10$, $\psi_2 = -0.15$, $\sigma_1 =  0.91$, $\sigma_2 = 0.87$,
  $z_* = 5.18, u_* = 0.874, v_*= 0.759, q_*= 0.267$,
  $EV=\left[0.000; 0.157; 2.064; -1.907\right]$,
  \item[3] $\psi_1 = 0.15$, $\psi_2 = 0.10$, $\sigma_1 = 1.18$, $\sigma_2 = 1.11$,
  $z_* = 7.56, u_* = 0.923, v_*= 0.818, q_*= 0.254$,
  $EV=\left[0.000; 0.174; 2.278; -2.104\right]$,
  \item[4] $\psi_1 = 0.10$, $\psi_2 = 0.15$, $\sigma_1 = 1.11$, $\sigma_2 = 1.18$,
  $z_* = 6.73,\;u_* = 0.933,\;v_*= 0.799,\;q_*= 0.259$,
  $EV=\left[0.000; 0.173; 1.872; -1.699\right]$,
\item[5] $\psi_1 = -0.15$, $\psi_2 = -0.10$, $\sigma_1 = 0.87$, $\sigma_2 = 0.91$,
  $z_* = 4.87, u_* = 0.884, v_*= 0.745, q_*= 0.271$,
  $EV=\left[0.000; 0.158; 1.773; -1.615\right]$.
\end{itemize}
Of course, there is another possible case where we have $\psi_1 =  \psi_2$, but this has already been analyzed by Chilarescu’s recent paper.
\section{Factor substitution and long-run equilibrium}
The idea of studying the effect of the elasticity of substitution on economic growth is not new, but the introduction of the concept of normalization by de La Grandville $(1989)$ and Klump and La Grandville $(2000)$, has allowed the creation of a methodology to gauge the sensitivity of the results to variations in the elasticity of substitution. The starting point of their attempt was to compare the performance of two economies, whose parameters are identical, except for the elasticity of substitution.

Before proceeding to the effective employment of this concept in our case, a few clarifications are necessary because some parameters require more attention. To define certain parameters, we use the derivative of the second $CES$ production function $wrt$ the effective physical capital and $wrt$ the effective human capital. As we have already pointed out, the education sector only produces human capital. Therefore, the first derivative of the function $y_2$ $wrt$ the variable $k(1-v)$ can be interpreted as the marginal variation in human capital created by the education sector for a unit increase in the effective physical capital employed in the education sector. Similarly, the same derivative $wrt$ the variable $h(1-u)$ can be interpreted as the marginal variation in human capital created by the education sector for a unit increase in human capital actually employed in the education sector. Therefore, the product $k(1-v)\frac{\partial y_2}{\partial\left[k(1-v)\right]}$ can be interpreted as the part of the human capital created by the education sector, due to the effective physical capital and, therefore, the relative share of physical capital, $\pi^k_2=\frac{k(1-v)}{y_2}\frac{\partial y_2}{\partial\left[k(1-v)\right]}$ is this fraction of the total human capital created in the education sector. Similarly, the product $h(1-u)\frac{\partial y_2}{\partial\left[h(1-u)\right]}$ can be interpreted as the part of the human capital created by the education sector, due to the effective human capital and, therefore, the relative share of human capital, $\pi^h_2=\frac{h(1-u)}{y_2} \frac{\partial y_2}{\partial\left[h(1-u)\right]}$ is this fraction of the total human capital created in the education sector. Of course we have $\pi^k_2 + \pi^h_2 = 1$. Also, we may define the  effective marginal rate of substitution for the sector of education, simply called marginal rate of substitution and will be defined as the ratio of the two first derivatives, $i.e.$ $m_2 = \frac{\frac{\partial y_2}{\partial\left[h(1-u)\right]}}{\frac{\partial y_2}{\partial\left[k(1-v)\right]}}$, and as we can see later this one is equal to $m_1$ and therefore we will denote by $m$ their common value.

The process of normalization of the $CES$ production function, $$y=A\left(\alpha k^\psi + 1-\alpha\right)^{\frac{1}{\psi}},$$ as was proposed by de La Grandville ($1989$), need some arbitrarily chosen baseline values, and those chosen by de La Grandville are: the capital-labor ratio $\bar{k}$, the per capita income $\bar{y}$ and the marginal rate of substitution $m$. The purpose of this procedure is to express the value of the parameters of the production function ($\alpha$, $A$ and if possible $\psi$), in terms of $\sigma$, $\bar{k}$, $\bar{y}$ and $m$. In order to use the same procedure, we need to modify the production function of the education sector $y_2$ (see the definition of $P_2$ in the equation \eqref{eqp1p2}), simply because the parameter $\theta$ is expressed as a function of parameters $\alpha_1$ and $\alpha_2$. Thus, using the relation \eqref{eqsr} we obtain:
\begin{equation}\label{eqpfy2}
y_2 =A_2h(1-u)\left[\alpha_2
\tau^{-\psi_2}w^{\psi_2}+1-\alpha_2\right]^{\frac{1}{\psi_2}}.
\end{equation}
With these clarifications, we can now proceed to the determination of these new concepts in order to better characterize the relationship between factor substitution and economic growth.
For given baseline values of capital-labor ratio $\bar{k}$, human capital $\bar{h}$, the fraction of physical capital allocated to the good sector $\bar{v}$, the fraction of labor allocated to the good sector $\bar{u}$, per-capita incomes $\bar{y}$ and the marginal rates of substitution $m_{1}$ and respectively $m_{2}$,
$$m_{1}=\frac{\frac{\partial y_1}{\partial (hu)}}{\frac{\partial y_1}{\partial (kv)}}=\frac{1-\alpha_1}{\alpha_1}\bar{w}^{1-\psi_1},\;
m_{2}=\frac{\frac{\partial y_2}{\partial \left[h(1-u)\right]}}{\frac{\partial y_2}{\partial \left[k(1-v)\right]}}=\frac{1-\alpha_2}{
\alpha_2}\left(\frac{\bar{w}}{\bar{\tau}}\right)^{1-\psi_2}=m_1,$$
the normalized $CES$ production functions become:
\begin{equation}\label{eqpfn}
\left\{
  \begin{array}{llll}
y_1(\sigma_1) =A_1\left(\sigma_1\right)\left(hu\right)\left[\alpha_1\left(\sigma_1\right)w^{\psi_1}
+1-\alpha_1\left(\sigma_1\right)\right]^{\frac{1}{\psi_1}},\\\\
y_2(\sigma_2) =A_2\left(\sigma_2\right)\left[h(1-u)\right]\left[\alpha_2\left(\sigma_2\right)
\tau^{-\psi_2}
w^{\psi_2}+1-\alpha_2\left(\sigma_2\right)\right]^{\frac{1}{\psi_2}},
\end{array}
  \right.
\end{equation}
where the efficiency and the distribution parameters are given by:
\begin{equation}\label{eqpalpha}
\alpha_1(\sigma_1) =\frac{\bar{w}^{1-\psi_1}}{\bar{w}^{1-\psi_1}+m},\;
\alpha_2(\sigma_2) =\frac{\left(\frac{\bar{w}}{\bar{\tau}}\right)^{1-\psi_2}}{\left(\frac{\bar{w}}{\bar{\tau}}\right)^{1-\psi_2}
+m},\;\bar{\tau}=
\frac{\bar{v}(1-\bar{u})}{\bar{u}(1-\bar{v})}
\end{equation}
\begin{equation}\label{eqpA}
A_1(\sigma_1) =\frac{\bar{y}_1}{\bar{h}\bar{u}}\left[\frac{\bar{w}^{1-\psi_1}
+m}{\bar{w}+m}\right]^{\frac{1}{\psi_1}},A_2(\sigma_2) =\frac{\bar{y}_2}{\bar{h}(1-\bar{u})}\left[\frac{\left(\frac{\bar{w}}{\bar{\tau}}\right)^{1-\psi_2}
+m}{\frac{\bar{w}}{\bar{\tau}}+m}\right]^{\frac{1}{\psi_2}}
\end{equation}
The physical capital shares are given by
\begin{equation}\label{eqpikh}
\left\{
  \begin{array}{llll}
\pi^k_{1}(\sigma_1) =\frac{\bar{w}^{1-\psi_1}w^{\psi_1}}{\bar{w}^{1-\psi_1}w^{\psi_1}+m},\;
\pi^k_{2}(\sigma_2) =\frac{\left(\frac{\bar{w}}{\bar{\tau}}\right)^{1-\psi_2}\left(\frac{{w}}{{\tau}}\right)^{\psi_2}}
{\left(\frac{\bar{w}}{\bar{\tau}}\right)^{1-\psi_2}\left(\frac{{w}}{{\tau}}\right)^{\psi_2}+m},\\\\
\bar{\pi}^k_{1} =\frac{\bar{w}}{\bar{w}+m},\;\;
\bar{\pi}^k_{2} =\frac{\frac{\bar{w}}{\bar{\tau}}}{\frac{\bar{w}}{\bar{\tau}}+m}
\;\mbox{if}\;w=\bar{w}\;\mbox{and}\;\tau=\bar{\tau}.\\
\end{array}
\right.
\end{equation}
As we can observe from the above relations, $\alpha_1$ and $A_1$ are expressed as functions of $\sigma_1$ and the baseline values ($\bar{w}$ and $m$), whereas $\alpha_2$ and $A_2$ are expressed as functions of $\sigma_2$ and the baseline values ($\bar{w}$, $m$ and $\bar{\tau}$). It should be also pointed out here that for $w \neq \bar{w}$, $\pi_1^k$ depends on $\sigma_1$  as well as on $w$, whereas the baseline share $\bar{\pi}_1^k$ is independent from both. Also, for $w \neq \bar{w}$ and $\tau \neq \bar{\tau}$, $\pi_2^k$ depends on $\sigma_2$, as well as on $w$ and $\tau$, whereas the baseline shares $\bar{\pi}_2^k$ is independent from them. As it is well-known, $w$ signify the ratio $kv/hu$. Therefore, we notice here that the ratio $w/\tau$ means in fact $k(1-v)/(h(1-u))$.
It is just a simply exercise to prove that, $\pi^h_{1} =1-\pi^k_{1}$, $\pi^h_{2} =1-\pi^k_{2}$.
Equations \eqref{eqpA} - \eqref{eqpikh}  enable us to reformulate the normalized functions \eqref{eqpfn} in the following simplified way
\begin{equation}\label{eqnfref}
\left\{
  \begin{array}{llll}
y_1=\bar{y}_1\frac{hu}{\bar{h}\bar{u}}\frac{w}{\bar{w}}\left[\frac{\bar{\pi}_1^k}
{\pi_1^k}\right]^{\frac{1}{\psi_1}}=\frac{\bar{y}_1}{\bar{k}\bar{v}}\left[\frac{\bar{\pi}_1^k}
{\pi_1^k}\right]^{\frac{1}{\psi_1}}{kv},\\\\
y_2=\bar{y}_2\frac{h(1-u)}{\bar{h}(1-\bar{u})}\frac{w}{\bar{w}}\frac{\bar{\tau}}{\tau}
\left[\frac{\bar{\pi}_2^k}{\pi_2^k}\right]^{\frac{1}{\psi_2}}=\frac{\bar{y}_2}{\bar{k}(1-\bar{v})}
\left[\frac{\bar{\pi}_2^k}{\pi_2^k}\right]^{\frac{1}{\psi_2}}{k(1-v)}.\\
\end{array}
\right.
\end{equation}
We are now able to highlight some interesting properties of the normalized production function. To obtain these results we use the following properties:
\begin{equation}\label{eqwwb}
\left(\frac{w}{\bar{w}}\right)^{\psi_1}=\frac{\pi^{k}_1\left(1-\bar{\pi}^k_1\right)}
{\bar{\pi}^k_1\left(1-\pi^{k}_1\right)},\;\left(\frac{w\bar{\tau}}{\bar{w}\tau}\right)^{\psi_2}
=\frac{\pi^{k}_2\left(1-\bar{\pi}^k_2\right)}
{\bar{\pi}^k_2\left(1-\pi^{k}_2\right)},
\end{equation}
that are obviously true, and can be proven by direct calculation. Also, since we have $\psi_1=\frac{\sigma_1-1}{\sigma_1}$ and $\psi_2=\frac{\sigma_2-1}{\sigma_2}$, it immediately follows that $$\frac{d y_1}{d \sigma_1}=\frac{\partial y_1}{\partial \psi_1}\frac{d \psi_1}{d \sigma_1} = \frac{\partial y_1}{\partial \psi_1}\frac{1}{\sigma_1^2}\;\mbox{and}\;\frac{d \pi_1}{d \sigma_1}=\frac{\partial \pi_1}{\partial \psi_1}\frac{d \psi_1}{d\sigma_1}=\frac{\partial\pi_1}{\partial \psi_1}\frac{1}{\sigma_1^2},$$
$$\frac{d y_2}{d \sigma_2}=\frac{\partial y_2}{\partial \psi_2}\frac{d \psi_2}{d \sigma_2} = \frac{\partial y_2}{\partial \psi_2}\frac{1}{\sigma_2^2}\;\mbox{and}\;\frac{d \pi_2}{d \sigma_2}=\frac{\partial \pi_2}{\partial \psi_2}\frac{d \psi_2}{d \sigma_2} = \frac{\partial \pi_2}{\partial \psi_2}\frac{1}{\sigma_2^2},$$
and therefore, the sign of the derivatives $wrt$ $\sigma_1 (\sigma_2)$ is the same with the sign of the derivatives $wrt$ $\psi_1 (\psi_2)$.
The equations describing the trajectories of physical and human capital become:
\begin{equation}\label{eqtrk}
\dot{k}=\left[\frac{\bar{y}_1}{\bar{k}\bar{v}}\left(\frac{\bar{\pi}_1^k}{\pi_1^k}\right)^{\frac{1}{\psi_1}}v
-\frac{c}{k}-\delta_k\right]k,
\end{equation}
\begin{equation}\label{eqtrh}
\dot{h}=\left[\frac{\bar{y}_2}{\bar{k}(1-\bar{v})}\left(\frac{\bar{\pi}_2^k}{\pi_2^k}\right)^{\frac{1}{\psi_2}}
\frac{w}{\tau}(1-u)-\delta_h\right]h.
\end{equation}
According to the equation \eqref{eqpikh}, it should be noted here again that,  $\pi_1^k$ depends on both $\sigma_1$ (via $\psi_1$) and $k$ and $\pi_2^k$ depends on both $\sigma_2$ (via $\psi_2$) and $h$. To be more precisely, $\pi_1^k$ is not only dependent on $\psi_1$ and $k$, but also on the effective physical $kv$, and effective human capital $hu$. Also, $\pi_2^k$ is not only dependent on $\psi_2$ and $h$, but also on the effective physical $k(1-v)$ and effective human capital $h(1-u)$.
We are now able to provide our first main theorem.
\begin{thm}
If two economies, each of them characterized by two sectors of production - good sector and education sector, are described by two distinct $CES$ production functions differing for each sector only by their elasticity of substitution, and sharing the same initially conditions, then at any stage of its development the economy with the higher elasticity of substitution will have a higher level of per capita income, a higher level of physical capital per capita and a higher level of human capital.
\end{thm}
\begin{prooft}
Let us first determine the derivatives of $\pi_1^k$ and $y_1$:
\begin{equation}\label{eqderpi1}
\frac{d\pi_1^k}{d\psi_1}=\pi_1^k\left(1-\pi_1^k\right)\ln\left(\frac{w}{\bar{w}}\right)
\left\{
\begin{array}{ll}
>0\;\forall\; w > \bar{w},\\
<0\;\forall\; w < \bar{w}.\\
   \end{array}
\right.
\end{equation}
As we can observe from the above equation, $\pi_1^k$ is an increasing function $wrt$ $\sigma_1$ if, the ratio $\frac{kv}{hu}$ is greater than the reference one $\frac{\bar{k}\bar{v}}{\bar{h}\bar{u}}$, otherwise, $\pi_1^k$ is a decreasing function $wrt$ $\sigma_1$.
$$\frac{dy_1}{d\psi_1}=\frac{\partial y_1}{\partial\psi_1}+\frac{\partial y_1}{\partial\pi_1}\frac{d\pi_1}{d\psi_1}=-\frac{1}{\psi_1^2}\frac{\bar{y}_1}{\bar{k}\bar{v}}\left[\frac{\bar{\pi}_1^k}
{\pi_1^k}\right]^{\frac{1}{\psi_1}}{kv}\left[\ln\left(\frac{\bar{\pi}_1^k}{\pi_1^k}\right)
+\left(1-\pi_1^k\right)\ln\left(\frac{w}{\bar{w}}\right)^{\psi_1}\right]$$
and using \eqref{eqwwb} we get
\begin{equation}\label{eqdery1}
\frac{dy_1}{d\psi_1}=-\frac{1}{\psi_1^2}y_1\left[\pi_1^k\ln\left(\frac{\bar{\pi}_1^k}{\pi_1^k}\right)
+\left(1-\pi_1^k\right)\ln\frac{1-\bar{\pi}^k_1}{1-\pi^{k}_1}\right]>0,
\end{equation}
since the strictly concavity of the logarithmic function implies $$\pi_1^k\ln\left(\frac{\bar{\pi}_1^k}{\pi_1^k}\right)
+\left(1-\pi_1^k\right)\ln\frac{1-\bar{\pi}^k_1}{1-\pi^{k}_1}<0.$$
Proceeding in the same way, we determine the derivatives of $\pi_2^k$ and $y_2$.
\begin{equation}\label{eqderpi2}
\frac{d\pi_2^k}{d\psi_2}=\pi_2^k\left(1-\pi_2^k\right)\ln\left(\frac{\frac{w}{\tau}}{\frac{\bar{w}}{\bar{\tau}}}
\right)>0\left\{
\begin{array}{ll}
>0\;\forall\; \frac{w}{\tau} > \frac{\bar{w}}{\bar{\tau}},\\
<0\;\forall\;  \frac{w}{\tau} < \frac{\bar{w}}{\bar{\tau}}.\\
   \end{array}
\right.
\end{equation}
$$\frac{dy_2}{d\psi_2}=-\frac{1}{\psi_2^2}\frac{\bar{y}_2}{\bar{k}(1-\bar{v})}
\left[\frac{\bar{\pi}_2^k}{\pi_2^k}\right]^{\frac{1}{\psi_2}}{kv}
\left[\ln\left(\frac{\bar{\pi}_2^k}{\pi_2^k}\right)
+\left(1-\pi_2^k\right)\ln\left(\frac{w\bar{\tau}}{\bar{w}\tau}\right)^{\psi_2}\right]$$
and using \eqref{eqwwb} we get
\begin{equation}\label{eqdery2}
\frac{dy_2}{d\psi_2}=-\frac{1}{\psi_2^2}y_2\left[\pi_2^k\ln\left(\frac{\bar{\pi}_2^k}{\pi_2^k}\right)
+\left(1-\pi_2^k\right)\ln\frac{1-\bar{\pi}^k_2}{1-\pi^{k}_2}\right]>0.
\end{equation}
From the equations \eqref{eqdery1} and \eqref{eqdery2} we deduce that $y_1$ and $y_2$ are strictly increasing functions of variables $\sigma_1$, respectively $\sigma_2$. Since, $\dot{k}$ is an increasing function of $\sigma_1$ (see equation \eqref{eqtrk}), then and at any point of time $t > 0$ the value of $k$ will be larger in the country with the largest elasticity of substitution. $\dot{h}$ is also an increasing function of $\sigma_2$ (see equation \eqref{eqtrh}), and therefore at any point of time $t > 0$ the value of $h$ will be larger in the country with the largest elasticity of substitution.  Since per capita income $y_1$ is itself
an increasing function of both $k$ and $\sigma_1$, it will always be larger in the country with the higher
elasticity of substitution. As we can observe, $y_2$ being an increasing function of both $h$ and $\sigma_2$, the quantity of human capital created will always be larger in the country with higher elasticity of substitution and thus the proof is completed.
\end{prooft}
\begin{rem}
As we pointed out in the introduction of this paper, all other papers assume that the elasticity of substitution is greater than one (first of all, in order to ensure the existence of an asymptotic positive growth rate - see the papers of Manuel Gomez and Palivos and Karagiannis). This restriction is not necessary here.
\end{rem}
Consider now the case of the steady state found in the previous section. As we have already pointed out, this stage of evolution of the economy always exists and is independent of the level of the elasticity of substitution $(\sigma)$, contrary to the results obtained by other papers, where this stage exists only if sigma is greater than one. The main purpose of this section is to analyse the effect of elasticity of substitution on the parameters characterizing the steady state. To be able to perform this analysis, we need to express these parameters in terms of variables $\pi$ and $\sigma$. Let us consider first the equation \eqref{eqsolr} describing the common growth rate along the equilibrium state.
$$r_*=\frac{1}{\varepsilon}\left\{\alpha_1A_1w_*^{\psi_1-1}\left[\alpha_1 w_*^{\psi_1}+1-\alpha_1\right]^{\frac{1}{\psi_1}-1}
-\rho-\delta_{k}\right\}.$$
Substituting $\alpha_1$ from equation \eqref{eqpalpha} and considering $\pi_1^k$ and $\bar{\pi}_1^k$ from equations \eqref{eqpikh} and $y_1$ from equation \eqref{eqnfref} and after some manipulations, the equation describing the common growth rate along the steady state becomes
\begin{equation}\label{eqrsst}
r_*\left(\sigma_1\right)=\frac{1}{\varepsilon}\left[\frac{\bar{y}_1}{\bar{k}\bar{v}}
\bar{\pi}_1^k\left(\frac{\bar{\pi}_{1}^k}{\pi_{1*}^k}\right)^{\frac{1-\psi_1}{\psi_1}}-\rho-\delta_k\right].
\end{equation}
The derivative of $r_*$ $wrt$ $\psi_1$ is given by
$$\frac{d r_*}{d\psi_1}=-\frac{\bar{y}_1}{\bar{k}\bar{v}}\frac{\bar{\pi}_1^k}{\varepsilon\psi_1^2}
\left(\frac{\bar{\pi}_{1}^k}{\pi_{1*}^k}\right)^{\frac{1-\psi_1}{\psi_1}}
\left\{\left[1-\left(1-\psi_1\right)\left(1-\pi_1^k\right)\right]\ln{\frac{\bar{\pi}_1^k}{\pi_1^k}}\right.\\\\$$
\begin{equation}\label{eqderpsi1}
\left.+\left(1-\psi_1\right)\left(1-\pi_1^k\right)\ln{\frac{1-\bar{\pi}_1^k}{1-\pi_1^k}}\right\}.
\end{equation}
Using the property of strict concavity of the logarithmic function we immediately obtain
$$\left[1-\left(1-\psi_1\right)\left(1-\pi_1^k\right)\right]\ln{\frac{\bar{\pi}_1^k}{\pi_1^k}}
+\left(1-\psi_1\right)\left(1-\pi_1^k\right)\ln{\frac{1-\bar{\pi}_1^k}{1-\pi_1^k}}
<\frac{\bar{\pi}_1^k-\pi_1^k}{\pi_1^k}$$
and we conclude that the common growth rate $r_*$ is an increasing function of the elasticity of substitution if the ratio $\frac{kv}{hu}$ is greater than the reference one $\frac{\bar{k}\bar{v}}{\bar{h}\bar{u}}$, that is if the share of physical capital is an increasing function.
\section{Numerical simulations and conclusions}
The main purpose of this section is to present some results of numerical simulations in order to confirm the theoretical aspects presented in the previous section and finally to give some conclusions. To do this we choose the following two alternatives. In the first one we consider two economies, $E1$ and $E2$,  both of them characterized by the following benchmark values: $A_1 = 1.05$, $A_2 = 0.20$, $\alpha_1 = 0.6$, $\alpha_2 = 0.8$, $\delta_k = 0.06$, $\delta_h = 0.05$, $\varepsilon = 2$, $\rho = 0.06$, $k_0 = 5.5$, $h_0 = 1$, $u_0 = 0.60$, $v_0 = 0.50$, but differing by the following elasticities of substitution:
\begin{enumerate}
  \item [E1:] $\psi_1 = 0.25$, $\psi_2 = -0.10$, $\sigma_1 =1.33$, $\sigma_2 = 0.91$.
  \item [E2:] $\psi_1 = 0.20$, $\psi_2 = -0.15$, $\sigma_1 =1.25$, $\sigma_2 = 0.87$.
\end{enumerate}
Applying the methodology described in the third section, we obtain the following results:
\begin{enumerate}
  \item [E1:] $k_* = 32.18$, $\pi^k_{1*} = 0.730$, $\pi^k_{2*} = 0.757$, $u_* = 0.8821$, $v_* = 0.8665$, $r_* = 0.1150$, $y_{1*} = 13.37$, $y_{2*} = 0.49$.
  \item [E2:] $k_* = 27.90$, $\pi^k_{1*} = 0.700$, $\pi^k_{2*} = 0.737$, $u_* = 0.8723$, $v_* = 0.8506$, $r_* = 0.1102$, $y_{1*} = 11.54$, $y_{2*} = 0.48$.
\end{enumerate}
As we can observe, the level of the parameters describing the steady-state for the economy with higher elasticity of substitution, are higher than those of the other economy.
\begin{center}
\includegraphics[width=6.0cm]{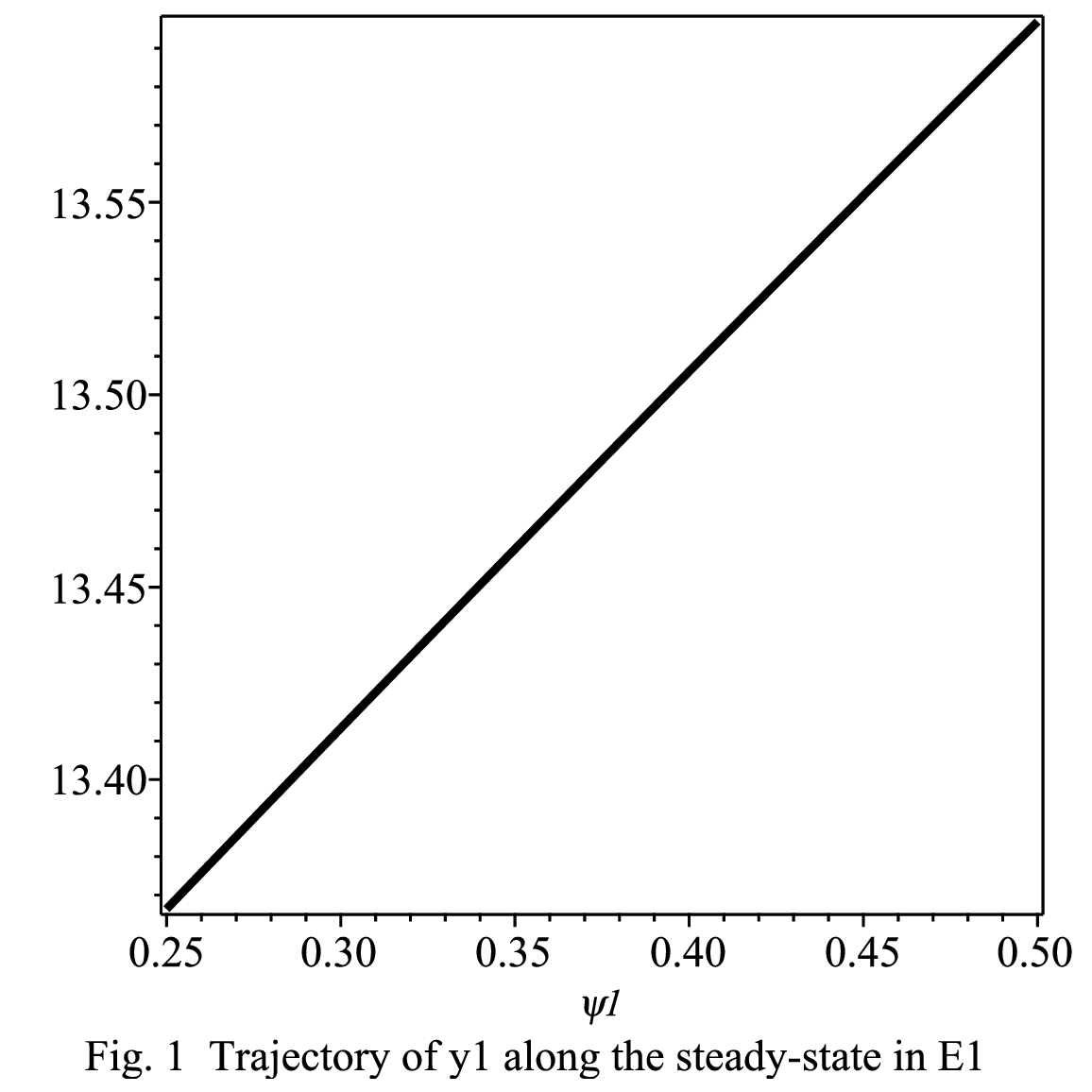}\;\;\;\;\;\includegraphics[width=6.0cm]{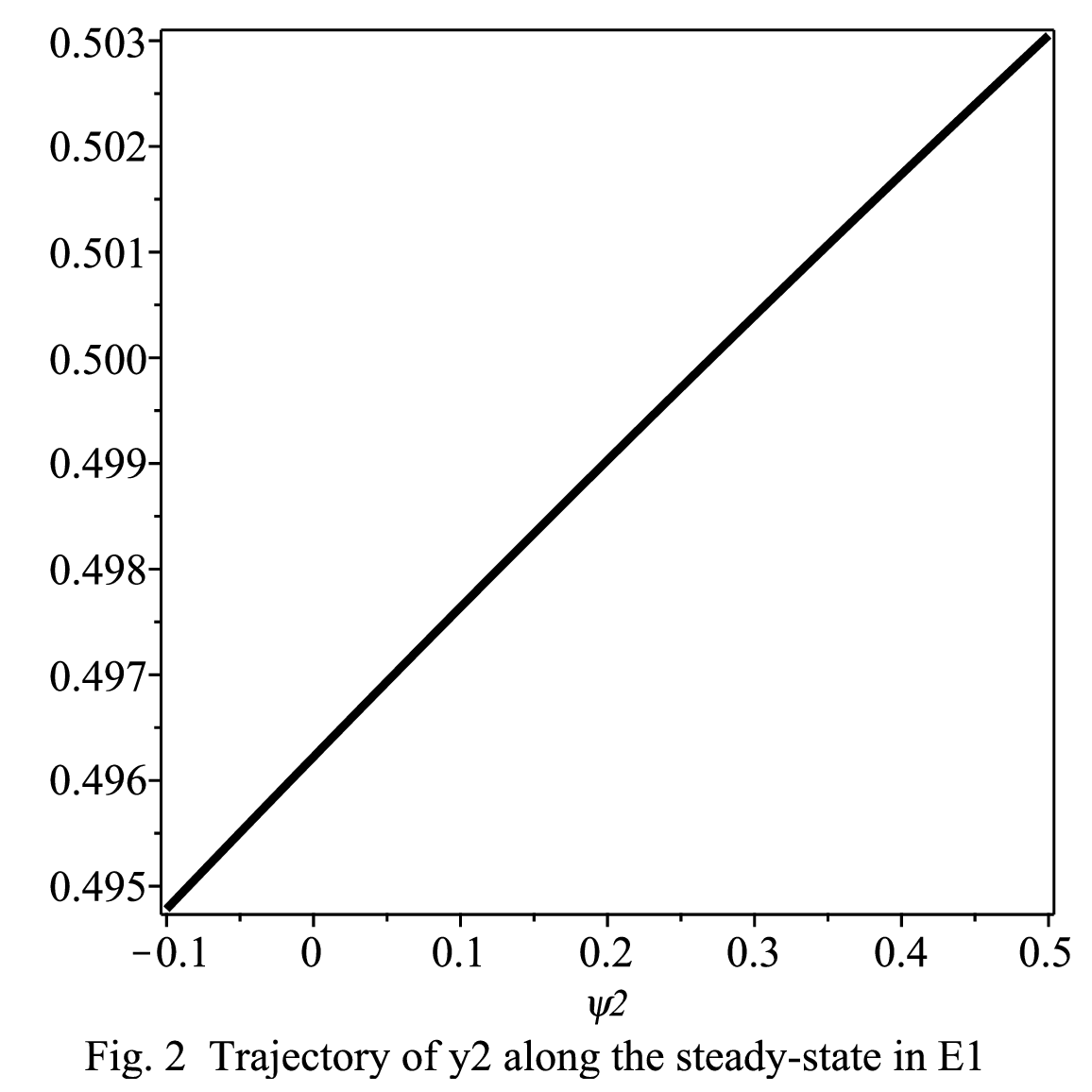}
\end{center}
\begin{center}
\includegraphics[width=6.0cm]{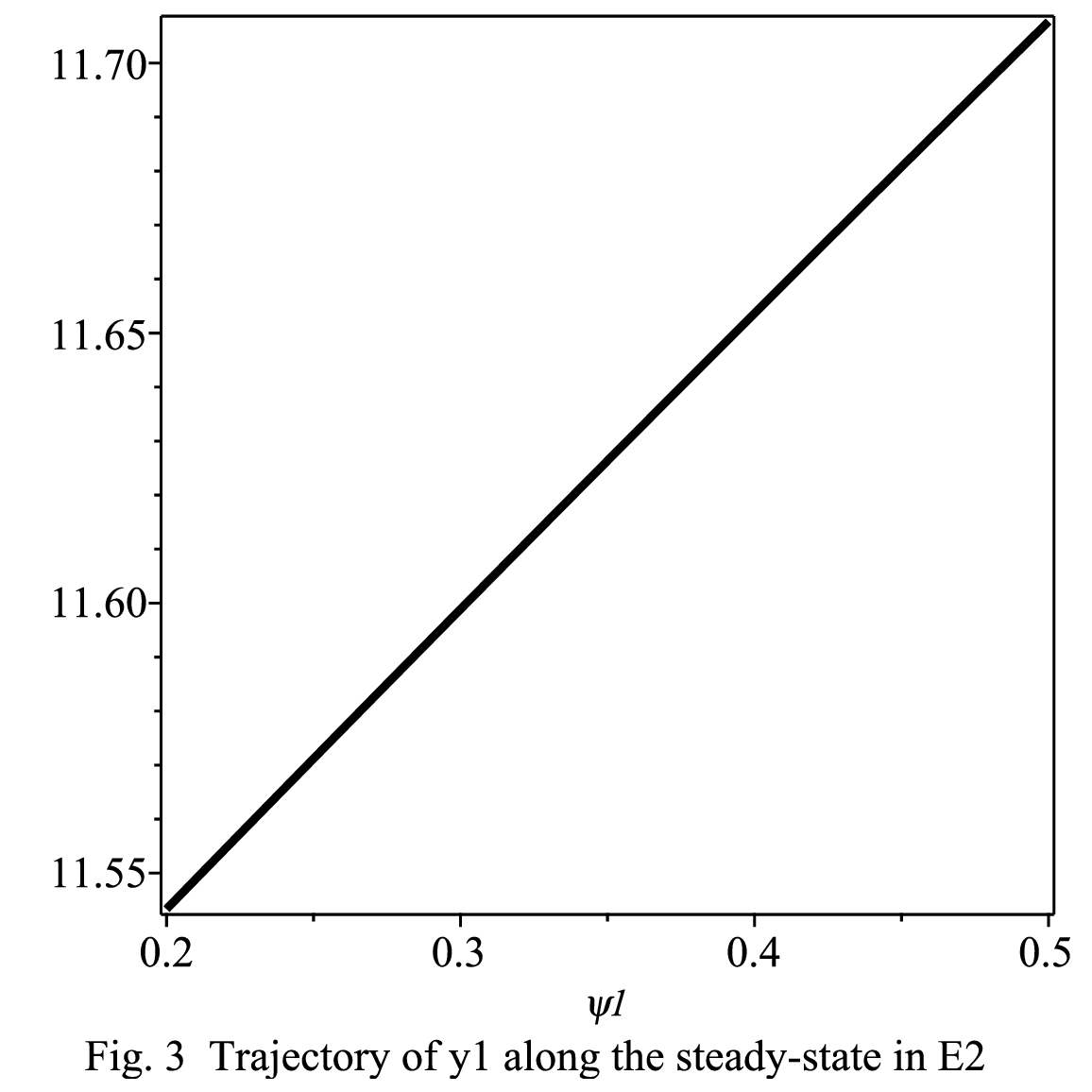}\;\;\;\;\;\includegraphics[width=6.0cm]{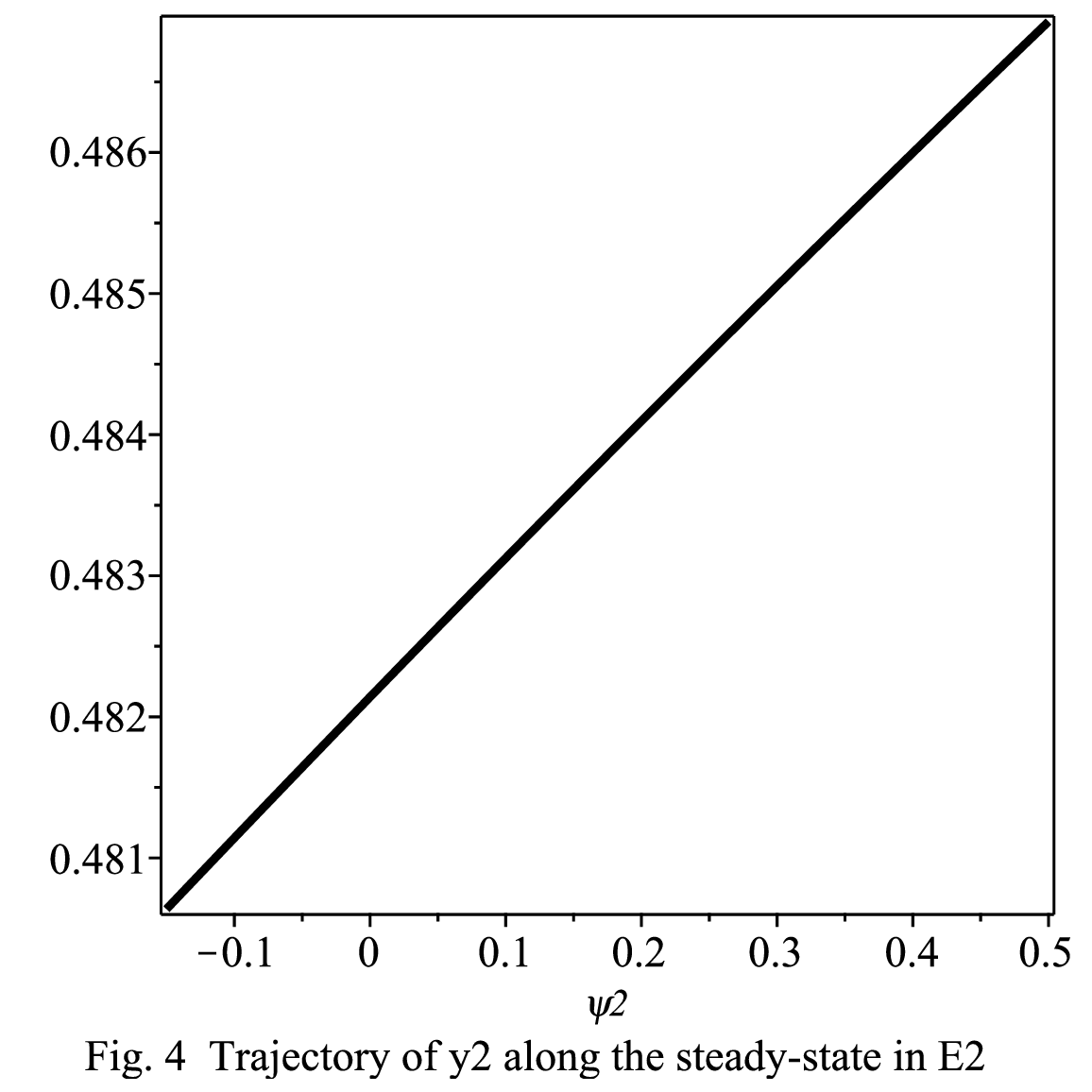}
\end{center}
The four above graphs show the evolution of the two production functions along the steady-state, for the two economies, as function of the elasticities of substitution and these graphs confirm that the two production functions are increasing functions of $\sigma_1$ respectively of $\sigma_2$. More than that, it can easily be observed that the evolution of the first economy (with a higher elasticity of substitution) is greater than that of the second. In the first alternative examined above, the elasticity of substitution of the good sector was considered above unity and that of the education sector was chosen below unity. The second alternative will examine the case where both elasticities are less than one. The benchmark values are the following, that is: $A_1 = 1.05$, $A_2 = 0.20$, $\alpha_1 = 0.6$, $\alpha_2 = 0.8$, $\delta_k = 0.06$, $\delta_h = 0.05$, $\varepsilon = 2$, $\rho = 0.06$, $k_0 = 5.0$, $h_0 = 1$, $u_0 = 0.50$, $v_0 = 0.30$, and the elasticities of substitution are as follows:
\begin{enumerate}
  \item [E1:] $\psi_1 = -0.10$, $\psi_2 = -0.15$, $\sigma_1 =0.91$, $\sigma_2 = 0.87$.
  \item [E2:] $\psi_1 = -0.15$, $\psi_2 = -0.20$, $\sigma_1 =0.87$, $\sigma_2 = 0.83$.
\end{enumerate}
Applying the methodology described in the third section, we obtain the following results:
\begin{enumerate}
  \item [E1:] $k_* = 15.56$, $\pi^k_{1*} = 0.563$, $\pi^k_{2*} = 0.739$, $u_* = 0.8739$, $v_* = 0.7593$, $r_* = 0.0976$, $y_{1*} = 6.61$, $y_{2*} = 0.44$.
  \item [E2:] $k_* = 14.70$, $\pi^k_{1*} = 0.547$, $\pi^k_{2*} = 0.720$, $u_* = 0.8647$, $v_* = 0.7499$, $r_* = 0.0949$, $y_{1*} = 6.25$, $y_{2*} = 0.43$.
\end{enumerate}
As we can observe, the level of the parameters describing the steady-state for the economy with higher elasticity of substitution, are higher than those of the other economy.
\begin{center}
\includegraphics[width=6.0cm]{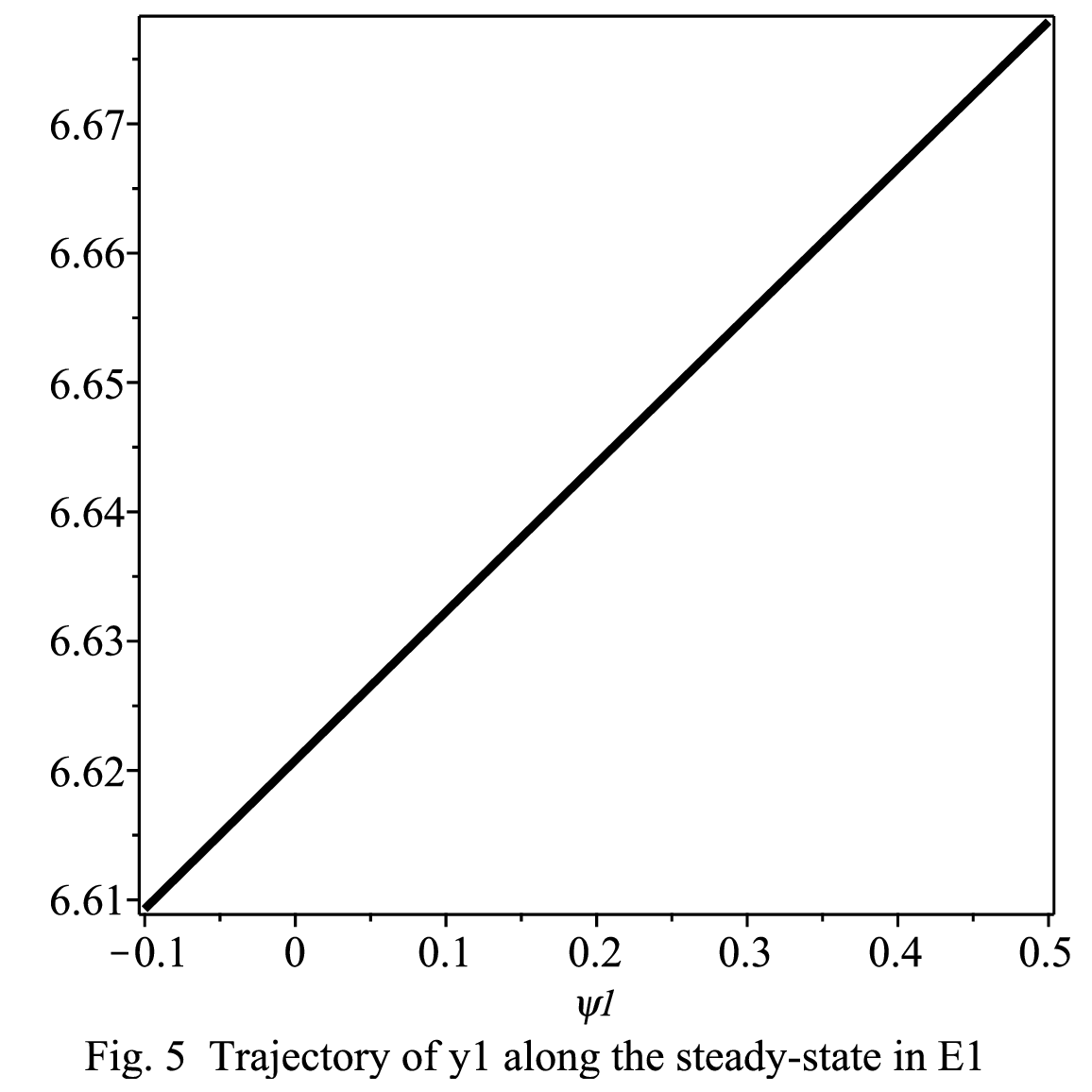}\;\;\;\;\;\includegraphics[width=6.0cm]{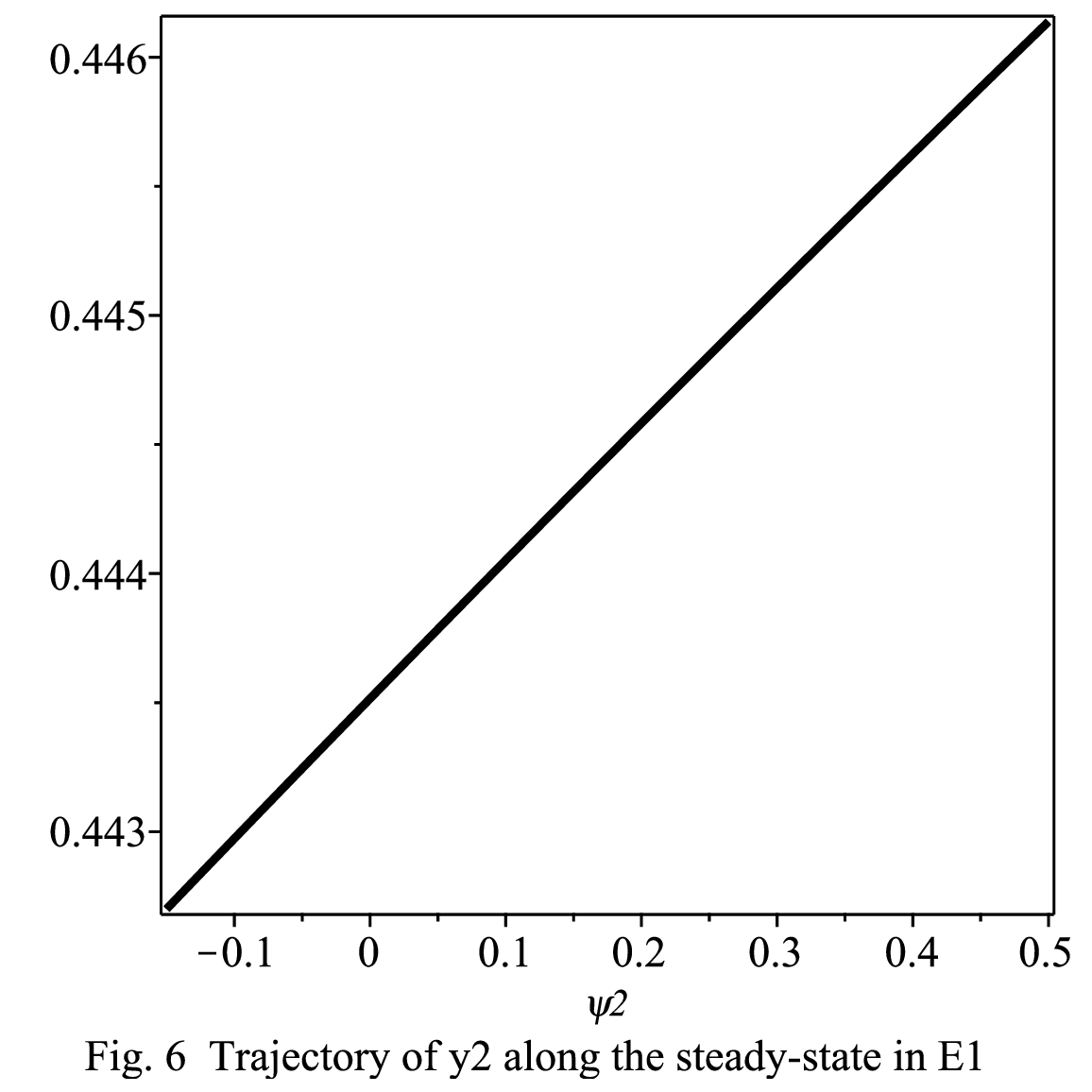}
\end{center}
\begin{center}
\includegraphics[width=6.0cm]{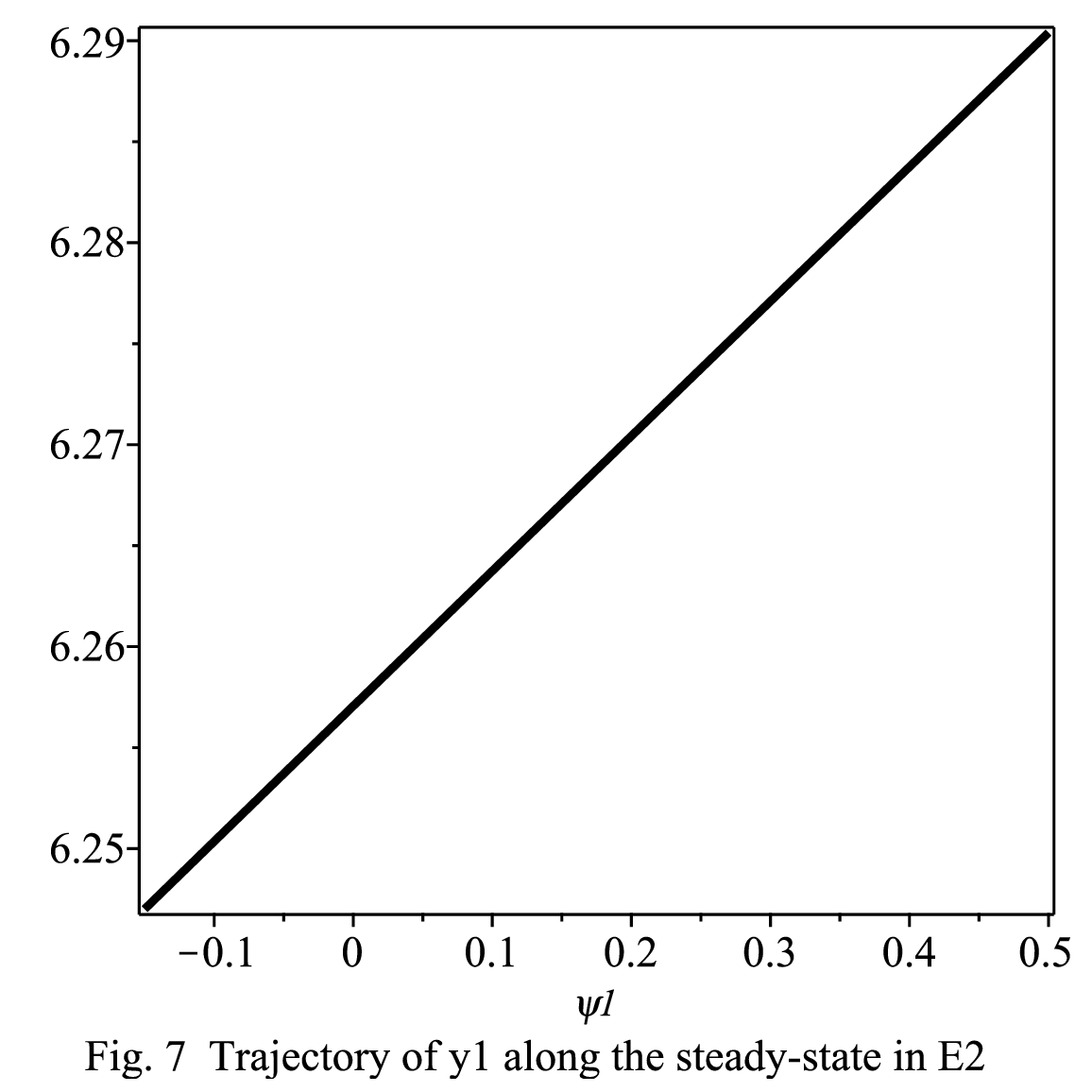}\;\;\;\;\;\includegraphics[width=6.0cm]{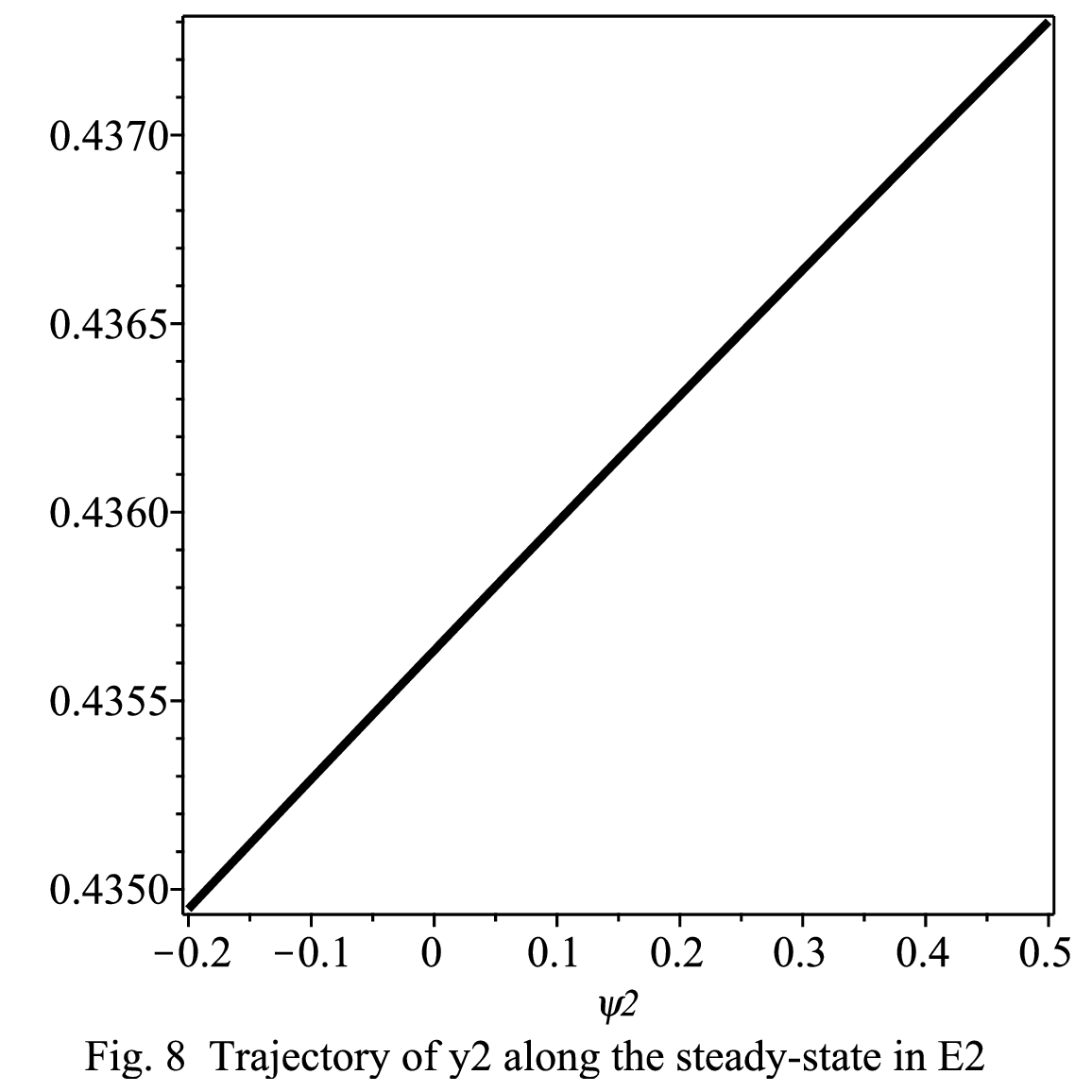}
\end{center}
The four above graphs show the evolution of the two production functions along the steady-state, for the two economies, as function of the elasticities of substitution and these graphs confirm that the two production functions are increasing functions of $\sigma_1$ respectively of $\sigma_2$. More than that, it can easily be observed that the evolution of the first economy (with a higher elasticity of substitution) is greater than that of the second. These results correspond to two economies with elasticities of substitution lower than one and the positive effect of substitution elasticity on economic growth is once again confirmed. This alternative is impossible to be considered in the case of an economy with one sector, or an economy with two sectors, but where physical capital not being present in the education sector (see the paper of Manuel Gomez).

% ----------------------------------------------------------------
\end{document}